\documentclass{ws-procs975x65}

\begin{document}

\title{A NEW LIMIT ON THE LIGHT SPEED ISOTROPY FROM THE GRAAL EXPERIMENT AT ESRF}

\author{V.G. GURZADYAN$^{1}$, V. BELLINI$^{2}$, M. BERETTA$^{3}$, J.-P. BOCQUET$^{4}$, A. D'ANGELO$^{5}$,\\ 
R.~DI SALVO$^{5}$,A. FANTINI$^{5}$, D. FRANCO$^{5}$, G. GERVINO$^{7}$, G. GIARDINA$^{8}$, F. GHIO$^{9}$,\\ 
B. GIROLAMI$^{9}$, A.~GIUSA$^{2}$, A. KASHIN$^{1}$, H.G. KHACHATRYAN$^{1}$, S. KNYAZYAN$^{1}$,\\ 
A. LAPIK$^{10}$, P. LEVI  SANDRI$^{3}$, A.~LLERES$^{4}$, F. MAMMOLITI$^{2}$, G. MANDAGLIO$^{8}$,\\ 
M. MANGANARO$^{8}$, A. MARGARIAN$^{1}$, S.~MEHRABYAN$^{1}$, R. MESSI$^{5}$,D. MORICCIANI$^{5}$,\\ 
V. NEDOREZOV$^{10}$, D. REBREYEND$^{4}$, G. RUSSO$^{2}$, N. RUDNEV$^{10}$, C.~SCHAERF$^{5}$,\\ 
M.-L. SPERDUTO$^{2}$, M.-C. SUTERA$^{2}$, A. TURINGE$^{10}$, V. VEGNA$^{5}$\\}

\address{$^{1}$Yerevan Physics Institute and Yerevan State University, 375036 Yerevan, Armenia\\
$^{2}$INFN sezione di Catania and Universita' di Catania, 95100 Catania, Italy\\
$^{3}$INFN Laboratori Nazionali di Frascati, 00044 Frascati, Italy\\
$^{4}$IN2P3, Laboratory for Subatomic Physics and Cosmology, 38026 Grenoble, France\\
$^{5}$INFN sezione di Roma TV and Universita' di Roma "Tor Vergata", 00133 Roma, Italy\\
$^{6}$INFN sezione di Genova and Universita' di Genova, 16146 Genova, Italy\\
$^{7}$INFN sezione di Torino  and Universita' di Torino, 10125 Torino, Italy\\
$^{8}$INFN sezione di Catania and Universita' di Messina, 98166 Messina, Italy\\
$^{9}$INFN sezione di Roma I and Istituto Superiore di Sanita', 00161 Roma, Italy\\
$^{10}$Institute for Nuclear Research, 117312 Moscow, Russia\\
}

\begin{abstract}
When the electrons stored in the ring of the European Synchrotron Radiation Facility (ESRF, Grenoble) scatter on a laser beam (Compton scattering in flight) the lower energy of the scattered electron spectra, the Compton Edge (CE), is given by the two body photon-electron relativistic kinematics and depends on the velocity of light. A precision measurement of the position of this CE as a function of the daily variations of the direction of the electron beam in an absolute reference frame provides a one-way test of Relativistic Kinematics and the isotropy of the velocity of light. The results of GRAAL-ESRF measurements improve the previously existing one-way limits, thus showing the efficiency of this method and the interest of further studies in this direction. 
\end{abstract}


\vspace{0.4in}

High precision tests of light speed isotropy are among the main studies to probe the limits of the Lorentz transformations and special relativity. Various experimental methods have been suggested and applied, each having its own interest, see \cite{Her,Ant,Ho} and refs therein.  The Compton Edge (CE) approach, i.e. the studies of the stability of the lowest energy of the scattered electrons after Compton scattering of relativistic electrons on monochromatic laser photons, also provide such a possibility, especially if we can perform many very stable measurements covering the daily rotations of the Earth. The parameters of the GRAAL beam-line\cite{Bo,Sc} at the European Synchrotron Radiation Facility (ESRF, Grenoble) enabled the performance of the measurements with the required precision. Below we represent some preliminary results of our 2008 measurements at the ESRF, while the detailed and complete report will be found in ref.[{\cite{Be}].  

The following aspects are peculiar to Compton Edge method \cite{GM}: (1) the Lorentz factor dependence of the Compton edge enables to reach high accuracy in the relative variations of CE with measurements within the available accelerator and laser parameters; (2) the choice of the inertial frame related to the dipole anisotropy, $\Delta T/T = 1.2\, 10^{-3}$, 
of the Cosmic Microwave Background (CMB) radiation, as the isotropy is a frame-dependent property; (3) the one-way character of the measurement and the fact that it depends only on the direction of the electron velocity; (4) the experimental limit obtained in this way can be used to constrain various models of Lorentz violation or extensions of special relativity.   

Among the hierarchy of motions \cite{RG} in which the Earth is participating and their corresponding frames, the frame with a null dipole anisotropy defines a 'rest' frame of CMB. The measurements with respect to that frame i.e. towards the CMB dipole's apex, $l=263.85^{\circ} \pm 0.1^{\circ},
b=48.25^{\circ}\pm 0.04^{\circ}$, are however, different from a Michelson-Morley type experiment, since the Michelson-Morley interferometer has two orthogonal arms with defined lengths (two-way experiment) while the Compton Edge measurement has only one direction and no typical length. Therefore CE provides more general, {\it one-way} constraints to the various extensions of special relativity (e.g. \cite{Ho}).   
     
The Compton edge of the scattered electrons is given by the formula\cite{mpla}
\begin{equation}
E_{CE}=\frac{4\gamma^2 E_l}{1+4 \gamma m_e E_l}=\frac{\gamma m_e X_{CE}}{A+X_{CE}},
\end{equation}
where $E_l$ is the energy of the laser photons, $\gamma$ is the Lorentz factor, $X_{CE}$ is the distance of the scattered electron from the circulating electron beam, $A=159.28\pm 0.2 mm$ is the constant of the magnetic dipole dispersion. GRAAL operated with three UV laser lines around $351\, nm$ and one Green line at $512\, nm$. The microstrip detector provided the positions of the scattered electrons, i.e. their energy in microstrip units with the given calibration and resolution of the system (see \cite{Bo,Sc,mpla}).  

Then, for ESRF 6.04 GeV electron beam ($\gamma = 11820$) one has
\begin{equation}
\Delta c/c= 0.7\,\, 10^{-8} \Delta X_{CE}/X_{CE}.
\end{equation}

The GRAAL-ESRF CE data acquired (non-continuously) during 1998-2002 had led to an upper limit for the light speed anisotropy \cite{mpla}: $\Delta c/c < 3\,\, 10^{-12}$.

In 2008 the data acquisition system has been enriched by a fast channel based on a VIRTEX-II DAQ which can operate at an acquisition rate up to 3 MHz allowing the full use of the scattered beam intensity of 800 kHz. The system will be described in refs.[\cite{Sc,Be}]. Two sessions of measurements have been performed in July and November. 
Some characteristics on the acquired data, with dates and number of points, standard deviations and $\chi^2$ for Gaussian distributions, separately for each set and together, are given in Table 1.

\begin{center}
%{\footnotesize
\begin{tabular}{crrcc}
\hline
 Data & Fragment (N) & Points & $\sigma$ & $\chi^2$ \\ 
\hline
\hline
 23.07.08 - 29.07.08 & 21-37 (15) & 14765 & $6.570\!\times\!10^{-5}$ & 0.97\\
 15.11.08 - 24.11.08 & 51-67 (17) & 18621 & $6.758\!\times\!10^{-5}$ & 2.16\\
 Total & 21-67 (32) & 33386 & $6.675\!\times\!10^{-5}$ & 1.89\\
\hline
\end{tabular}%}
\end{center}

The 2008 data (here and below jointly of July and November) distribution as function of the solar 24-hour day and sidereal days are shown in Fig. 1ab. 

\begin{figure}[thb]%
\begin{center}
\psfig{file=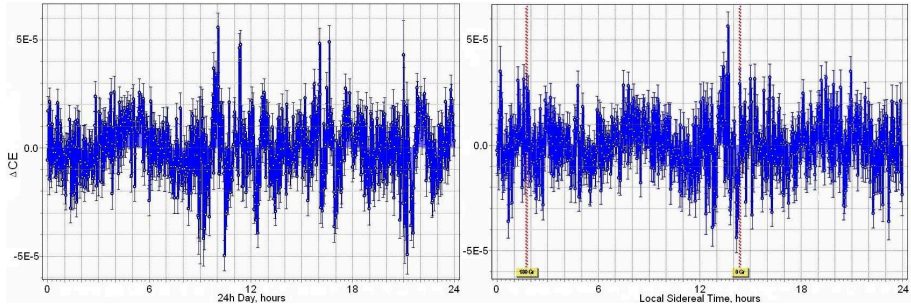,width=0.95\textwidth}
\caption{CE data vs 24-hour solar (a) and sidereal (b) days for ESRF 2008 measurements.}
\end{center}
\end{figure}

Figures 2 and 3 show the distribution of the CE data vs the angle between the GRAAL beam and the CMB dipole apex and
the corresponding hours of day, respectively (cf. \cite{mpla}). CE 2-4 $\sigma$ variations are visible both at angle and hour correlations, as well
as in their Fourier spectra. The Fourier spectra of the CE vs the angle and the hours are shown in Fig. 4; the abscissa axes are given both in multipoles and degrees, hours. 

\begin{figure}[thb]%
\begin{center}
\psfig{file=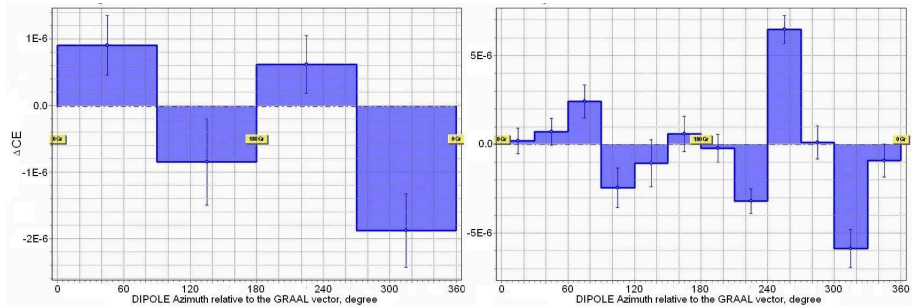,width=0.95\textwidth}
\caption{CE vs angle between the GRAAL beam and the CMB dipole apex, averaged within  $90^{\circ}$ (a) and  $30^{\circ}$ (b) intervals.}
\end{center}
\end{figure}

\begin{figure}[thb]%
\begin{center}
\psfig{file=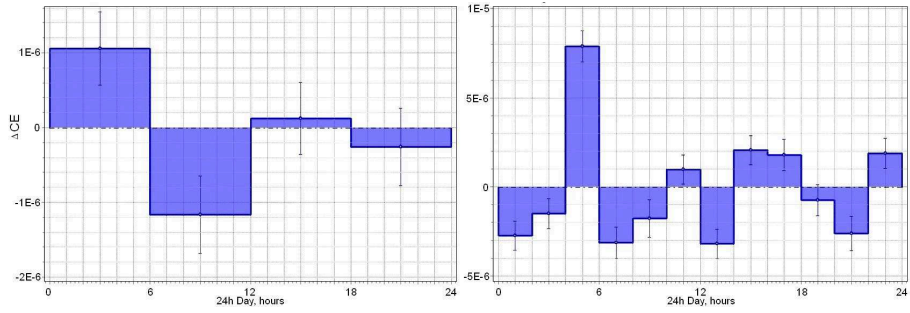,width=0.95\textwidth}
\caption{CE vs time, averaged within 6 (a) and 2-hour (b) intervals.}
\end{center}
\end{figure}

\begin{figure}[thb]%
\begin{center}
\psfig{file=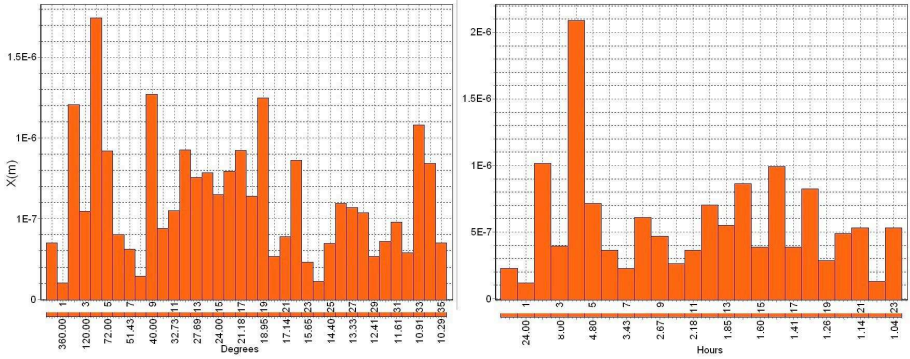,width=0.95\textwidth}
\caption{Fourier spectra of CE dependence vs the GRAAL-dipole angle (a) and hours (b), respectively.}
\end{center}
\end{figure}

The interpretation of the nature of the 2-4$\sigma$ CE variations in Figs.2-4 (at early morning hours) requires further studies of the systematic daily variations in the GRAAL-ESRF environment. They correspond to a maximum displacement of the tagging detector of 250 $nm$ or an unexplained change in the temperature of the tagging box of 0.01 C. 
The Fourier spectrum in Fig. 4b clearly shows that there is no evidence of the 24 hours cycle that one would expect from the rotation of the Earth.

From the above analysis a one-way isotropy limit for the light speed is obtained
\begin{equation}
\Delta c/c \lesssim 1.0\,\, 10^{-14}.
\end{equation}
The new, fast acquisition system, has allowed a reduction of our previous upper limit by more than two orders of magnitude. \vspace{0.1in}

{\it Acknowledgements.} It is a pleasure to thank the Director and Staff of the ESRF Accelerator and Source Division for a stable and reliable operation of the storage ring, G. Nobili for his competent and dedicated technical support with the experimental apparatus and S. Angeloni for his expert advice and contribution with the Virtex and associated hardware and software.

\bibliographystyle{ws-procs975x65}
\end{document}